\documentclass[11pt]{article} 
\newcommand{\be}{\begin{equation}}
\newcommand{\ee}{\end{equation}}
\newcommand{\bea}{\begin{eqnarray}}
\newcommand{\eea}{\end{eqnarray}}
\newcommand{\sn}{{\rm sn}}
\newcommand{\ds}{{\rm ds}}
\newcommand{\cs}{{\rm cs}}
\newcommand{\ns}{{\rm ns}}
\newcommand{\dn}{{\rm dn}}
\newcommand{\cn}{{\rm cn}}
\newcommand{\sech}{{\rm sech}}

\begin{document}
\vspace{.5in} 
\begin{center} 
{\LARGE{\bf Novel PT-invariant Solutions For a Large Number 
of Real Nonlinear Equations}}
\end{center} 

\vspace{.3in}
\begin{center} 
{\LARGE{\bf Avinash Khare}} \\ 
{Physics Department, Savitribai Phule Pune University \\
Pune, India 411007}
\end{center} 

\begin{center} 
{\LARGE{\bf Avadh Saxena}} \\ 
{Theoretical Division and Center for Nonlinear Studies, Los
Alamos National Laboratory, Los Alamos, NM 87545, USA}
\end{center} 

\vspace{.9in}
{\bf {Abstract:}}  

For a large number of real nonlinear equations, either continuous 
or discrete, integrable or nonintegrable, we show that whenever a real 
nonlinear equation admits a solution in terms of $\sech x$, it also admits 
solutions in terms of the PT-invariant combinations $\sech x \pm i \tanh x$. 
Further, for a number of real nonlinear equations we show that whenever 
a nonlinear equation admits a solution in terms $\sech^2 x$, it also admits 
solutions in terms of the PT-invariant combinations $\sech^2 x \pm i \sech x 
\tanh x$. Besides, we show that similar results are also true in the periodic 
case involving Jacobi elliptic functions.

\newpage 
  
\section{Introduction} 

Nonlinear equations are playing an increasingly important role in several areas 
of science in general and physics in particular. One of the major problems with 
these equations is the lack of a superposition principle. 
It is thus necessary to explicitly obtain more and more solutions of a 
given nonlinear equation. Thus if we can find some general results about
the existence of solutions to a nonlinear equation, that would be invaluable. 
In this context it is worth 
recalling that some time ago we \cite{ks} had shown (through a number of
examples) that if a nonlinear equation admits a periodic solution in terms of 
Jacobi elliptic functions $\dn(x,m)$ and $\cn(x,m)$, then it will also admit 
solutions in terms of $\dn(x,m)\pm\cn(x,m)$, where $m$ is the modulus of the 
elliptic function. 
Further, in the same paper \cite{ks}, we also showed (again through 
several examples) that if a nonlinear equation admits a solution in terms of 
$\dn^2(x,m)$, then it will also admit solutions in terms of $\dn^2(x,m) \pm 
\cn(x,m) \dn(x,m)$. 

The purpose of this paper is to propose general results about the
existence of new solutions to real nonlinear equations, integrable or
nonintegrable, continuous or discrete through the idea of PT-symmetry.
It may be noted here that in the last 15-20 years the idea of PT 
symmetry \cite{ben} has given us new insight. In quantum mechanics  
it has been shown that even if Hamiltonian is not hermitian but if it
is PT-invariant, then the energy eigenvalues are still real in case 
PT symmetry is not broken spontaneously. Further, there is tremendous
growth in the number of studies of open systems which are specially
balanced by PT symmetry \cite{sch,ben1,peng} in several 
PT-invariant open systems bearing both loss and gain, one has obtained  
soliton solutions and they have been shown to be stable within certain 
parameter range \cite{mak,pan,jes}. 

In this paper we highlight one more novel
aspect of PT-symmetry.  In particular, we obtain 
new PT-invariant solutions through a general principle. We show, 
through several examples, that whenever a real nonlinear
equation, either continuous or discrete, integrable or nonintegrable, 
admits a solution in terms of $\sech x$, then it will necessarily
also admit solutions in terms of the PT-invariant combinations
$\sech x \pm i \tanh x$. We also generalize these results to the periodic
case and show that whenever a nonlinear equation
admits a solution in terms of $\dn(x,m)$ [or $\cn(x,m)]$, then it will 
necessarily also admit solutions in terms of the PT-invariant combinations 
$\dn (x,m)\pm i\sqrt{m}\sn (x,m)$ [or $\cn(x,m) \pm i \sn(x,m)$]. 

Further, we show, through several examples, that whenever a real nonlinear
equation admits a solution in terms of $\sech^2 x$, then it will also admit
solutions in terms of $\sech^2 x \pm i \sech x \tanh x$. We also generalize 
these results to the periodic case and show that whenever a real nonlinear
equation admits a solution in terms of $\dn^2(x,m)$, then it will necessarily
also admit solutions in terms of $\dn^2(x,m)\pm im\sn(x,m)\cn(x,m)$ as 
well as $\dn^2(x,m)\pm i \sqrt{m}\sn(x,m)\dn(x,m)$.

\section{Solutions in Terms of $\sech x\pm i \tanh x$ as well as 
Their Periodic Generalization}

We now discuss four examples, two from continuum field theories and
two from the discrete case where $\sech x$ is a known solution and in all 
the four cases we obtain new PT-invariant
solutions in terms of $\sech x \pm i \tanh x$ and also periodic PT-invariant
solutions in terms of $\dn (x,m) \pm i \sn (x,m)$ as well as $\cn (x,m) \pm i 
\sn (x,m)$.

\subsection{$\phi^4$ Field Theory}

The $\phi^4$ field theory arises in several areas of physics including 
second order phase transitions.
The field equation for the $\phi^2-\phi^4$ field theory is given by
\be\label{1.14}
\phi_{xx} = a\phi+b\phi^3\,,
\ee
In case $b < 0$, one of the well known solution to this equation is 
\be\label{1.15}
\phi = A \sech[\beta x]\,,
\ee
provided
\be\label{1.16}
b A^2 = - 2\beta^2\,,~~a = \beta^2\,.
\ee

Remarkably, even 
\be\label{1.17}
\phi = A \sech(\beta x) \pm i B\tanh(\beta x)\, 
\ee
is an exact PT-invariant solution of Eq. (\ref{1.14}) provided
\be\label{1.18}
B = \pm A\,,~~2b A^2 = -\beta^2\,,~~ a = -(1/2)\beta^2\,.
\ee

Further, as we now show,  such PT-invariant solutions also exist in the 
periodic case. Let us first note that one of the exact, 
periodic solution to the $\phi^4$  Eq. (\ref{1.14}) is
\be\label{1.19}
\phi = A\dn(\beta x,m)\,,
\ee
provided
\be\label{1.20}
b A^2 = - 2\beta^2\,,~~a = (2-m)\beta^2\, . 
\ee

Further, the same model (\ref{1.14}) is known to admit another periodic
solution
\be\label{1.21}
\phi = A \sqrt{m} \cn(\beta x,m)\,,
\ee
provided
\be\label{1.22}
b A^2 = - 2\beta^2\,,~~a = (2m-1)\beta^2\,.
\ee

Remarkably, we find that the same model also admits the PT-invariant 
periodic solution
\be\label{1.23}
\phi = A \dn(\beta x,m)+ i B\sqrt{m} \sn(\beta x,m)\,,
\ee
provided
\be\label{1.24}
B = \pm A\,,~~2b A^2 =-\beta^2\,,~~a = -\frac{2m-1}{2}\beta^2 . 
\ee

Further, the same model also admits another PT-invariant solution
\be\label{1.25}
\phi = A \sqrt{m} \cn[\beta x,m]+ i B\sqrt{m} \sn[\beta x,m]\,,
\ee
provided
\be\label{1.26}
B = \pm A\,,~~2b A^2 =-\beta^2\,,~~a = -\frac{2-m}{2}\beta^2\,.
\ee

\subsection{mKdV Equation}

We first discuss the celebrated mKdV equation
\be\label{1.1}
u_t+u_{xxx}+ 6 u^2 u_{x} =0\,,
\ee
which is a well known integrable equation having application in several 
areas \cite{dj}. It is well known that 
\be\label{1.2}
u = A \sech[\beta(x-vt)\, 
\ee
is an exact solution of Eq. (\ref{1.1}) provided 
\be\label{1.3}
A^2=\beta^2\,,~v = \beta^2\,.
\ee

Remarkably,  even
\be\label{1.4}
u = A \sech[\beta(x-vt)] \pm i B \tanh[\beta(x-vt)]\, 
\ee
is also an exact PT-invariant solution to the mKdV Eq. (\ref{1.1}) provided
\be\label{1.5}
B = \pm A\,,~~A^2 = 4\beta^2\,,~~v= -(1/2) \beta^2\,.
\ee

Even more remarkable, such PT-invariant solutions also exist in the 
periodic case. For example, it is well known that one of the exact, 
periodic solution to the mKdV Eq. (\ref{1.1}) is \cite{as}
\be\label{1.6}
u = A \dn[\beta(x-vt),m]\,,
\ee
provided
\be\label{1.7}
A^2 =  \beta^2\,,~~v= (2-m) \beta^2\,.
\ee
Another periodic solution to the mKdV Eq. (\ref{1.1}) is
\be\label{1.8}
u = A \sqrt{m} \cn[\beta(x-vt),m]\,,
\ee
provided
\be\label{1.9}
A^2 =  \beta^2\,,~~v= (2m-1) \beta^2\,.
\ee

Remarkably, even 
\be\label{1.10}
u = A\dn[\beta(x-vt),m]+i B \sqrt{m} \sn[\beta(x-vt),m]\, 
\ee
is an exact PT-invariant solution to the mKdV Eq. (\ref{1.1}) provided
\be\label{1.11}
B = \pm A\,,~~A^2 = 4\beta^2\,,~~v= -\frac{(2m-1)}{2}\beta^2\,.
\ee
We thus have two new periodic solutions of mKdV Eq. (\ref{1.1}) 
depending on whether $B=A$ or $B=-A$. 

Further, even 
\be\label{1.12}
u = A\sqrt{m} \cn[\beta(x-vt),m]+i B \sqrt{m} \sn[\beta(x-vt),m]\,,
\ee
is an exact PT-invariant solution of the mKdV Eq. (\ref{1.1}) provided
\be\label{1.13}
B = \pm A\,,~~A^2 = 4\beta^2\,,~~v= -\frac{(2-m)}{2}\beta^2\,.
\ee

\subsection{Discrete $\phi^4$ Equation}

We now discuss two discrete models and show that both these models also admit
PT-invariant solutions. Let us first consider the discrete $\phi^4$ equation  
\be\label{1.27}
\frac{1}{h^2}[\phi_{n+1}+\phi_{n-1}-2\phi_n]+ a \phi_n 
-\frac{\lambda}{2} \phi_{n}^{2}[\phi_{n+1}+\phi_{n-1}] = 0\,.
\ee

It is well known that the Eq. (\ref{1.27}) admits an exact solution
\be\label{1.28}
\phi_n = A \sech(\beta n)\,,
\ee
provided
\be\label{1.29}
A^2  = - \frac{2 \sinh^2(\beta)}{h^2 \lambda}\,,~~
a h^2 = -4 \sinh^2(\beta/2)\,.
\ee

Remarkably, the same model also admits a PT-invariant periodic solution
\be\label{1.30}
\phi_n = A \sech(\beta n) \pm i B\tanh(\beta n)\,,
\ee
provided
\be\label{1.31}
B = \pm A\,,~~A^2   = - \frac{2\tanh^2 (\beta/2)}{h^2 \lambda}\,,
~~a h^2 = 2\tanh^2(\beta/2) \,.
\ee

Besides, the same model also has novel, PT-invariant periodic solutions.
Let us first note that a well known exact periodic solution to the 
Eq. (\ref{1.27}) is
\be\label{1.32}
\phi_n = A\dn(\beta n,m)\,,
\ee
provided
\be\label{1.33}
A^2 \cs^2(\beta, m) = - \frac{2}{h^2 \lambda}\,,~~
a h^2 = 2\left[1-\frac{dn(\beta,m)}{\cn^2(\beta,m)}\right]\, , 
\ee
where $\cs(x,m)=\cn(x,m)/\sn(x,m)$.  Further, the same model (\ref{1.27}) is 
known to admit another periodic solution
\be\label{1.34}
\phi_n = A \sqrt{m} \cn(\beta n,m]\,,
\ee
provided
\be\label{1.35}
A^2 \ds^2(\beta, m) = - \frac{2}{h^2 \lambda}\,,~~
a h^2 = 2\left[1-\frac{cn(\beta,m)}{\dn^2(\beta,m)}\right]\, , 
\ee
where $\ds(x,m)=\dn(x,m)/\sn(x,m)$. 

We find that the same model also admits the PT-invariant 
periodic solution
\be\label{1.36}
\phi_n = A\dn(\beta n,m) +iB \sqrt{m} \sn(\beta n,m)\,,
\ee
provided
\be\label{1.37}
B = \pm A\,,~~A^2 [\cs(\beta, m)+\ns(\beta,m)]^2  = - \frac{2}{h^2 \lambda}\,,
~~a h^2 = 2\left[1-\frac{2\dn(\beta,m)}{1+\cn(\beta,m)}\right]\,.
\ee
Further, the same model also admits another PT-invariant solution
\be\label{1.38}
\phi_n = A \sqrt{m} \cn(\beta n,m)+ i B\sqrt{m} \sn(\beta n,m)\,,
\ee
provided
\be\label{1.39}
B = \pm A\,,~~A^2 [\ds(\beta, m)+\ns(\beta,m)]^2  = - \frac{2}{h^2 \lambda}\,,
~~a h^2 = 2\left[1-\frac{2\cn(\beta,m)}{1+\dn(\beta,m)}\right]\,.
\ee
While deriving results in this and the next subsection, we have made use of
several not so well known identities satisfied by the Jacobi elliptic functions
\cite{kls}.

\subsection{Discrete mKdV Equation}

Let us consider the discrete mKdV equation  
\be\label{1.40}
\frac{du_n}{dt}+  \alpha (u_{n+1} -u_{n-1})
+ \lambda u_{n}^{2}(u_{n+1} -u_{n-1}) =0\,. 
\ee
It is well known that this model has an exact 
hyperbolic soliton solution 
\be\label{1.41}
u_n = A \sech[\beta (n-vt)]\,,
\ee
provided
\be\label{1.42}
\lambda A^2  = \alpha \sinh^2(\beta)\,,~~
\beta v = 2 \alpha \sinh(\beta)\,.
\ee

We find that this model also admits the PT-invariant solution
\be\label{1.43}
u_n = A \sech(\beta n) \pm i B\tanh(\beta n)\,,
\ee
provided
\be\label{1.44}
B = \pm A\,,~~\lambda A^2 = \alpha \tanh^2(\beta/2)\,,
~~\beta v  = 4 \alpha \tanh(\beta/2) \,.
\ee

We find that this model also admits exact PT-invariant periodic solutions.
Let us first note that the well known periodic solution to Eq. (\ref{1.40})
is
\be\label{1.45}
u_n = A\dn[\beta (n-vt),m]\,,
\ee
provided
\be\label{1.46}
\lambda A^2 \cs^2(\beta, m) = \alpha\,,
~~\beta v = \frac{2\alpha}{\cs(\beta,m)}\,.
\ee

Further, the same model (\ref{1.40}) is known to admit another periodic
solution
\be\label{1.47}
u_n = A \sqrt{m} \cn[\beta (n-vt),m]\,,
\ee
provided
\be\label{1.48}
\lambda A^2 \ds^2(\beta, m) = \alpha\,,~~
\beta v = \frac{2\alpha}{\ds(\beta,m)}\,.
\ee

We now show that the same model also admits a PT-invariant 
periodic solution
\be\label{1.49}
u_n = A\dn[\beta (n-vt),m] +iB \sqrt{m} \sn[\beta (n-vt),m]\,,
\ee
provided
\be\label{1.50}
B = \pm A\,,~~\lambda A^2 [\cs(\beta, m)+\ns(\beta,m)]^2  = \alpha\,,~~
\beta v = \frac{4\alpha \sn(\beta,m)}{1+\cn(\beta,m)}\, , 
\ee
where $\ns(x,m)=1/\sn(x,m)$.  Further, the same model also admits another 
PT-invariant solution
\be\label{1.51}
u_n = A \sqrt{m} \cn(\beta n,m)+ i B\sqrt{m} \sn(\beta n,m)\,,
\ee
provided
\be\label{1.52}
B = \pm A\,,~~\lambda A^2 [\ds(\beta, m)+\ns(\beta,m)]^2  = \alpha\,,~~
\beta v = \frac{4\alpha \sn(\beta,m)}{1+\dn(\beta,m)}\, . 
\ee

\section{Solutions in Terms of $\sech^2 x\pm i \sech x\tanh x$ as well as 
Their Periodic Generalization}

We now discuss two examples where $\sech^2 x$ is a known solution and in both 
the cases we obtain new PT-invariant
solutions in terms of $\sech^2 x \pm i \sech x\tanh x$ and also 
PT-invariant periodic 
solutions in terms of $\dn^2(x,m) \pm i m\sn(x,m) \cn(x,m)$ as well as 
$\dn^2(x,m) \pm i \sqrt{m} \sn(x,m) \dn(x,m)$.

\subsection{KdV Equation}

We first discuss the celebrated KdV equation
\be\label{2.1}
u_t+u_{xxx}- 6 u u_{x} =0\,,
\ee
which is a well known integrable equation having application in several areas
including shallow water waves \cite{dj}. It is also well known that it admits 
the soliton solution  
\be\label{2.2}
u = A \sech^2(x-vt)\,,
\ee
provided $A= -2\beta^2\,,~v = 4 \beta^2$.
Remarkably, it also admits a PT-invariant solution
\be\label{2.3}
u = A \sech^2(x-vt) + i B \sech(x-vt) \tanh(x-vt)\,,
\ee
provided
\be\label{2.4}
B = \pm A\,,~A = -\beta^2\,,~v = \beta^2\,. 
\ee

We now show that KdV equation also admits periodic PT-invariant solutions.
It is well known that one of the exact, periodic solution to the 
KdV Eq. (\ref{2.1}) is
\be\label{2.5}
u = A \dn^2[\beta(x-vt),m]\,,
\ee
provided
\be\label{2.6}
A =  -2\beta^2\,,~~v= 4 (2-m) \beta^2\,.
\ee
Remarkably, even 
\be\label{2.7}
u = A\dn^2[\beta(x-vt),m]+i B m \sn[\beta(x-vt),m] \cn[\beta(x-vt),m]\,,
\ee
is an exact solution of the KdV Eq. (\ref{2.1}) provided
\be\label{2.8}
B = \pm A\,,~~A = -\beta^2\,,~~v= -(2-m)\beta^2\,.
\ee
We thus have two new periodic solutions of  the KdV Eq. (53) 
depending on whether $B=A$ or $B=-A$. 

Remarkably, there is another PT-invariant solution to the same KdV equation 
\be\label{2.9}
u = A\dn^2[\beta(x-vt+\delta_1),m]+
i B \sqrt{m} \sn[\beta(x-vt),m] \dn[\beta(x-vt),m]\,,
\ee
provided
\be\label{2.10}
B = \pm A\,,~~A = -\beta^2\,,~~v= (5-4m)\beta^2\,.
\ee

Few remarks are in order at this stage.

\begin{enumerate}

\item It is well known that the hyperbolic potential 
-$2\beta^2 \sech^2(\beta x)$ which
is a solution of the KdV equation, is a 
reflectionless potential. We then predict that the potentials 
-$\beta^2 \sech^2(\beta x)\pm i \sech(\beta x) \tanh(x)$ must also be 
reflectionless potentials. 

\item It is well known that the periodic potential 
-$2\beta^2 \dn^2(\beta x,m)$ which
is a solution of the KdV equation, has
precisely one band gap. We then predict that the potentials 
-$\beta^2 \dn^2(\beta x,m)\pm i m \beta^2 \sn(\beta x,m) \cn(\beta x,m)$ 
as well as the potentials 
$\beta^2 \dn^2(\beta x,m)\pm i \sqrt{m} \beta^2 \sn(\beta x,m) \dn(\beta x,m)$ 
must also have precisely one band gap.

\end{enumerate}

\subsection{$\phi^3$ Field Theory}

This field theory arises in the context of third order phase 
transitions \cite{phi3} and is also relevant to tachyon condensation 
\cite{tachyon}.  
The field equation for the $\phi^2-\phi^3$ field theory is given by
\be\label{2.11}
\phi_{xx} = a\phi+b\phi^2\,,
\ee
which is known to admit an exact solution
\be\label{2.12}
\phi = A \sech^2(\beta x)+B\,,
\ee
provided
\be\label{2.13}
A = -\frac{3a}{2b}\,,~~\beta^2 = \frac{a}{4}\,,~~B = 0\,.
\ee
Remarkably, Eq. (\ref{2.11}) also admits a PT-invariant solution 
\be\label{2.14}
\phi = A \sech^2[\beta(x)] \pm i D \sech[\beta(x)] \tanh[\beta(x)] +B\,,
\ee
provided
\be\label{2.15}
D = \pm A\,,~~A = -\frac{3a}{b}\,,~~\beta^2 = a\,,~~B = 0\,.
\ee

Further the model also admits PT-invariant periodic solutions. Let us
first note that the model (\ref{2.11}) also admits the periodic solution
\be\label{2.16}
\phi = A\dn^2[\beta(x),m]+B\,,
\ee
provided
\be\label{2.17}
A=-\frac{3a}{2b\sqrt{1-m+m^2}}\,,~~\beta^2=\frac{a}{4\sqrt{1-m+m^2}}\,,~~
B=\frac{a[2-m-\sqrt{1-m+m^2}]}{2b\sqrt{1-m+m^2}}\,.
\ee
It is easy to show that the same model also admits a PT-invariant periodic 
solution
\be\label{2.18}
\phi = A \dn^2[\beta(x),m]+ i D\sqrt{m} \cn[\beta(x),m]
\dn[\beta(x),m]+B\,,
\ee
provided
\be\label{2.19}
D= \pm A\,,~~A =-\frac{3a}{b\sqrt{16-16 m+m^2}}\,,~~\beta^2
=\frac{a}{\sqrt{16-16 m+m^2}}\,,~~
B=\frac{a[2-m-\sqrt{16-16 m+m^2}]}{2b\sqrt{16-16 m+m^2}}\,.
\ee
Further, the same model also admits another PT-invariant periodic solution 
\be\label{2.20}
\phi = A \dn^2[\beta(x),m]+ i D\sqrt{m} \sn[\beta(x),m]
\dn[\beta(x+c),m]+B\,,
\ee
provided
\be\label{2.21}
D= \pm A\,,~~A =-\frac{3a}{b\sqrt{1-16 m+16 m^2}}\,,~~\beta^2
=\frac{a}{\sqrt{1-16 m+16 m^2}}\,,~~
B=\frac{a[5-4m-\sqrt{1-16 m+16 m^2}]}{2b\sqrt{1-16 m+16 m^2}}\,.
\ee

\section{PT-Invariant Solutions in Three Coupled  models}

We now consider three different coupled models and show that in all these 
cases one has PT-invariant solutions for all the coupled fields.

\subsection{Coupled $\phi^4$ Model}

We first consider a coupled $\phi^4$ model 
\be\label{3.1}
\phi_{xx} = 2a_1 \phi + 4 b_1 \phi^3 +2\gamma \phi \psi^2\,,
\ee
\be\label{3.2}
\psi_{xx} = 2a_2 \psi + 4 b_1 \psi^3 +2\gamma \psi \phi^2\,,
\ee
and show that even in this case,
PT-invariant solutions are allowed in both the fields.
 
It is well known that this coupled system admits the solution \cite{ks1} 
\be\label{3.3}
\phi = A \sech(\beta x)\,,~~\psi = D \sech(\beta x)\,,
\ee
provided
\be\label{3.4}
2b_1 A^2 +\gamma D^2 = - \beta^2 = 2b_2 D^2 + \gamma A^2\,,
~~a_1 = a_2 = \frac{\beta^2}{2}\,.
\ee
Remarkably, the coupled model also admits the PT-invariant solution
\bea\label{3.5}
&&\phi = A \sech(\beta x)+ i B \tanh(\beta x)\,, \nonumber \\
&&\psi = D \sech(\beta x)+ i F \tanh(\beta x)\,,
\eea
provided
\bea\label{3.6}
&&B= \pm A\,,~~F = \pm D\,,~~a_1 = a_2 
= -\frac{\beta^2}{4}\,, \nonumber \\
&&4(2b_1 A^2+\gamma D^2) =-\beta^2 = 4(2b_2 D^2+\gamma A^2)\,.
\eea
Note that the signs of $B = \pm A$ and $F = \pm D$ are correlated.

This coupled model also admits PT-invariant periodic solutions. Let us first 
note that one of the well known periodic solution to the coupled 
Eq. (\ref{3.1}) is
\be\label{3.7}
\phi = A\dn[\beta x,m]\,,~~\psi = D\dn[\beta x, m]\,,
\ee
provided
\be\label{3.8}
2b_1 A^2 +\gamma D^2 = - \beta^2 = 2b_2 D^2 + \gamma A^2\,,
~~a_1 = a_2 = \frac{(2-m)\beta^2}{2}\,.
\ee
Further, the same coupled model is known to admit another periodic
solution
\be\label{3.9}
\phi = A \sqrt{m} \cn[\beta x,m]\,,~~\psi = D \sqrt{m} \cn[\beta x, m]\,,
\ee
provided
\be\label{3.10}
2b_1 A^2 +\gamma D^2 = - \beta^2 = 2b_2 D^2 + \gamma A^2\,,
~~a_1 = a_2 = \frac{(2m-1)\beta^2}{2}\,.
\ee

Remarkably, we find that the same coupled model also admits a PT-invariant
periodic solution
\bea\label{3.11}
&&\phi = A \dn[\beta x,m]+ i B\sqrt{m} \sn[\beta x,m]\,, \nonumber \\
&&\psi = D \dn[\beta x,m]+ i F\sqrt{m} \sn[\beta x,m]\,,
\eea
provided
\bea\label{3.12}
&&B= \pm A\,,~~F = \pm D\,,~~a_1 = a_2 
= -\frac{(4m-3)\beta^2}{4}\,, \nonumber \\
&&4(2b_1 A^2+\gamma D^2) =-\beta^2 = 4(2b_2 D^2+\gamma A^2)\,.
\eea
Note that the signs of $B = \pm A$ and $F = \pm D$ are correlated.

Further, the same model also admits another PT-invariant periodic solution
\bea\label{3.13}
&&\phi = A \sqrt{m} \cn[\beta x,m]+ i B\sqrt{m} \sn[\beta x,m]\,, \nonumber \\
&&\psi = D \sqrt{m}  \cn[\beta x,m]+ i F\sqrt{m} \sn[\beta x,m]\,,
\eea
provided
\bea\label{3.14}
&&B= \pm A\,,~~F = \pm D\,,~~a_1 = a_2 
= -\frac{(4-3m)\beta^2}{4}\,, \nonumber \\
&&4(2b_1 A^2+\gamma D^2) =-\beta^2 = 4(2b_2 D^2+\gamma A^2)\,.
\eea
Note that the signs of $B = \pm A$ and $F = \pm D$ are correlated.

\subsection{Coupled KdV Equations}

We now discuss the coupled KdV model which has also received some 
attention in the literature \cite{zhou} and show that even in this case,
PT-invariant solutions exist in both the coupled fields. 

The coupled KdV equations are
\bea\label{3.15}
&&u_t + \alpha u u_x + \eta v v_{x} + u_{xxx} = 0\,, \nonumber \\
&&v_t + \delta u v_x + v_{xxx} = 0\,.
\eea
One of the well known solution to the coupled Eqs. (\ref{3.15}) is \cite{zhou} 
\be\label{3.16}
u = A\sech^2[\beta(x-ct)]\,,~~v = D\sech^2[\beta(x-ct)]\,,
\ee
provided
\be\label{3.17}
\delta A = 12 \beta^2\,,~~\eta D^2 = (\delta - \alpha)A^2\,,~~
c = 4 \beta^2\,.
\ee
Remarkably, the same coupled model also admits the hyperbolic PT-invariant 
solution
\bea\label{3.18}
&&u = A\sech^2[\beta(x-ct)]+iB \tanh[\beta(x-ct)]\sech[\beta(x-ct)]\,,
\nonumber \\
&&v = D\sech^2[\beta(x-ct)]+iF \tanh[\beta(x-ct)]\sech[\beta(x-ct)]\,,
\eea
provided
\be\label{3.19}
B = \pm A\,,~~F = \pm D\,,~~\delta A = 6 \beta^2\,,~~\eta D^2 
= (\delta - \alpha)A^2\,,~~c = \beta^2\,.
\ee

This discussion is easily generalized to the periodic case. In particular, 
it is easy to check that the coupled Eqs. (\ref{3.15}) have the periodic
solution
\be\label{3.20}
u = A\dn^2[\beta(x-ct),m]\,,~~v = D\dn^2[\beta(x-ct),m]\,,
\ee
provided
\be\label{3.21}
\delta A = 12 \beta^2\,,~~\eta D^2 = (\delta - \alpha)A^2\,,~~
c = 4(2-m) \beta^2\,.
\ee
Remarkably, the same model also admits a PT-invariant periodic solution
\bea\label{3.22}
&&u = A\dn^2[\beta(x-ct),m]+iB m \sn[\beta(x-ct),m]\cn[\beta(x-ct),m]\,,
\nonumber \\
&&v = D\dn^2[\beta(x-ct),m]+iF m \sn[\beta(x-ct),m]\cn[\beta(x-ct),m]\,,
\eea
provided
\be\label{3.33}
B = \pm A\,,~~F = \pm D\,,~~\delta A = 6 \beta^2\,,~~\eta D^2 
= (\delta - \alpha)A^2\,,~~c = (2-m) \beta^2\,.
\ee
Note that the signs of $B = \pm A$ and $F = \pm D$ are correlated.
Further, the same model also admits another PT-invariant periodic solution
\bea\label{3.34}
&&u = A\dn^2[\beta(x-ct),m]+iB \sqrt{m} \sn[\beta(x-ct),m]\dn[\beta(x-ct),m]\,,
\nonumber \\
&&v = D\dn^2[\beta(x-ct),m]+iF \sqrt{m} \sn[\beta(x-ct),m]\dn[\beta(x-ct),m]\,,
\eea
provided
\be\label{3.35}
B = \pm A\,,~~F = \pm D\,,~~\delta A = 6 \beta^2\,,~~\eta D^2 
= (\delta - \alpha)A^2\,,~~c = (2m-1) \beta^2\,.
\ee
Note that the signs of $B = \pm A$ and $F = \pm D$ are correlated.

\subsection{Coupled KdV-mKdV Model}

Finally we consider a coupled KdV-mKdV model  
\bea\label{3.36}
&&u_t + u_{xxx} + 6u u_x+2\alpha u v v_x = 0\,, \nonumber \\
&&v_t + v_{xxx} + 6v^2 v_x+\gamma  v u_x = 0\,,
\eea
and show that in this case too we have PT-invariant solutions of the form
$\sech^2 x \pm i \sech x \tanh x$ and $\sech x \pm i \tanh x$ in KdV and
mKdV fields, $u$ and $v$, respectively.  

It is easy to check that
\be\label{3.37}
u = A\sech^2[\beta(x-ct)]+G\,,~~v = D\sech[\beta(x-ct)]\,,
\ee
is an exact solution of the coupled Eqs. (\ref{3.36}) provided
\be\label{3.38}
12 D^2 +4\gamma A = 12 \beta^2 = 6A +\alpha D^2\,,~~
c = \beta^2\,,~~G = -\frac{A}{4}\,.
\ee
Remarkably, the same model also admits a PT-invariant solution
\bea\label{3.39}
&&u = A\sech^2[\beta(x-ct)]+iB \tanh[\beta(x-ct)]\sech[\beta(x-ct)]\,,
\nonumber \\
&&v = D\sech[\beta(x-ct)]+iF \tanh[\beta(x-ct)]\,,
\eea
provided
\be\label{3.40}
B = \pm A\,,~~F = \pm D\,,~~12D^2 +2\gamma A = 3 \beta^2 = 3A +\alpha D^2\,,~~
c = -\frac{1}{2} \beta^2\,,~~G = -\frac{A}{4}\,.
\ee

We now show that the same model also has PT-invariant periodic solutions. 
Let us first note that
\be\label{3.41}
u = A\dn^2[\beta(x-ct),m]+G\,,~~v = D\dn[\beta(x-ct),m]\,,
\ee
is an exact solution of the coupled Eqs. (\ref{3.36}) provided
\be\label{3.42}
12 D^2 +4\gamma A = 12 \beta^2 = 6A +\alpha D^2\,,~~
c = (2-m) \beta^2\,,~~G = -\frac{(2-m)A}{4}\,.
\ee
It is easy to check that the same model also admits a PT-invariant solution
\bea\label{3.43}
&&u = A\dn^2[\beta(x-ct),m]+iB \sqrt{m} \sn[\beta(x-ct),m]\dn[\beta(x-ct),m]
+G\,, \nonumber \\
&&v = D\dn[\beta(x-ct),m]+iF m \sn[\beta(x-ct),m]\,,
\eea
provided
\be\label{3.44}
B = \pm A\,,~~F = \pm D\,,~~12D^2 +2\gamma A = 3 \beta^2 = 3A +\alpha D^2\,,~~
c = -\frac{(2m-1)}{2} \beta^2\,,~~G = -\frac{(3-2m)A}{4}\,.
\ee
Note that the signs of $B = \pm A$ and $F = \pm D$ are correlated.
Further, the same model also admits another PT-invariant solution
\bea\label{3.45}
&&u = A\dn^2[\beta(x-ct),m]+iB m \sn[\beta(x-ct),m]\cn[\beta(x-ct),m]\,,
\nonumber \\
&&v = D \sqrt{m} \cn[\beta(x-ct),m]
+iF \sqrt{m} \sn[\beta(x-ct),m]\dn[\beta(x-ct),m]\,,
\eea
provided
\be\label{3.46}
B = \pm A\,,~~F = \pm D\,,~~12D^2 +2\gamma A = 3 \beta^2 = 3A +\alpha D^2\,,~~
c = -\frac{(2-m)}{2} \beta^2\,,~~G = -\frac{(2-m)A}{4}\,.
\ee
Note that the signs of $B = \pm A$ and $F = \pm D$ are correlated.

\section{Summary and Conclusions}

In this paper we have shown through several examples that whenever a
real nonlinear equation admits solution in terms of $\sech x$ (or $\sech^2 x$), 
then the same model also admits solutions in terms of $\sech x \pm i \tanh x$
(or $\sech^2 x \pm i \sech x \tanh x$). Further, we have also shown that such
PT-invariant solutions also exist in the corresponding periodic case involving 
Jacobi elliptic functions. 

The obvious open  question is whether these results are true in general. It
would be nice if one can prove this in general, both in the hyperbolic as well 
as in the periodic case. In the absence of a general proof, it is worthwhile 
looking at more and more examples and see if this observation is true in 
general or if there are some exceptions. The other question is:  
What could be the deeper underlying reason because of which such 
solutions exist? Another question is about the significance of
such solutions for a real nonlinear equation. In this context we would like
to remark that the symmetry of solutions of a nonlinear equation need not
be the same as that of the nonlinear equation but could be less. Normally, 
the complex solutions of a real nonlinear equation are not of 
relevance. However, being $PT$ invariant complex solutions, we believe 
they could have some physical significance.
One pointer in this direction is the fact that for both the KdV and the 
mKdV equations, which are integrable equations, we have checked that the first 
3 constants of motion for the PT-invariant complex solutions 
of both the KdV and the mKdV equations are in fact real but have 
different values then those for the usual hyperbolic solution 
(and we suspect that in fact all the constants of motion would be real and would be different than those for the real hyperbolic solution) thereby suggesting 
that such solutions could be physically interesting. Thus it would be 
worthwhile studying the stability of such PT-invariant solutions. 
That may shed some light on the possible significance of such solutions.

We hope to address some of these issues in the near future.

\section{Acknowledgments} 

One of us (AK) is grateful to B. Dey and P. Durga Nandini for stimulating 
discussions and to INSA for the award of INSA senior Scientist position at 
Savitribai Phule Pune University. This work was supported in part by the 
U.S. Department of Energy.


\begin{thebibliography}{99}

\bibitem{ks} A. Khare and A. Saxena, Phys. Lett. {\bf A 377} (2013) 2761; 
J. Math. Phys. {\bf 55} (2014) 032701.

\bibitem{ben} See for example, C.M. Bender, Rep. Prog. Phys. {\bf 70} 
(2007) 947 and references therein. 

\bibitem{sch} See for example, J. Schindler, Z. Lin, MC Zheng, FM Ellis
and T. Kottos, Phys. Rev. {\bf A 84} (2011) 040101; J. Schindler, Z. Lin, 
JM Lee, H. Ramezani, FM Ellis and T. Kottos, J. Phys. A: Math. Theor.
{\bf 45} (2012) 444029.  

\bibitem{ben1} CM Bender, B. Berntson, D. Parker and E. Samuel, Am. J. 
Phys. {\bf 81} (2013) 173.

\bibitem{peng} B. Peng, SK \"Ozdemir, F. Lei, F. Monifi, M. Gianfreda, 
GL Long, S. Fan, F. Nori, CM Bender and L. Young, arXiv: 1308.4564. 

\bibitem{mak} KG Makris, R. El-Ganainy, DN Christodoulides and ZH Musslimani,
Int. J. Theor. Phys. {\bf 50} (2011) 1019.

\bibitem{pan} PG Kevrekidis, J. Cuevas-Maraver, A. Saxena, F. Cooper, 
A. Khare, arXiv:1507.03211. 

\bibitem{jes} J. Cuevas-Maraver, PG Kevrekidis, A. Saxena, F. Cooper, 
A. Khare, A. Comech and CM Bender, arXiv:1508.00852. 

\bibitem{dj} See for example, {\it Solitons: An Introduction} by
P.G. Drazin and R.S. Johnson, (Cambridge Univ. Press. 1989) and 
references therein. 

\bibitem{as} See for example, M. Abromowitz and I.A. Stegun, Handbook
of Mathematical Functions (Dover Publications, New York, 2010).


\bibitem{kls} A. Khare and U.P. Sukhatme, J. Math. Phys. {\bf 43} (2002) 
3798; A. Khare, A. Lakshminarayan and U.P. Sukhatme, 
ibid {\bf 44} (2003) 1822; Pramana {\bf 62} (2004) 1201.

\bibitem{phi3} P. Kumar, D. Hall, and R.G. Goodrich, Phys. Rev. Lett.
{\bf 82} (1999) 4532; P. Kumar and A. Saxena, Phil. Mag. {\bf B 82} 
(2002) 1201.

\bibitem{tachyon} B. Zwiebach, J. High Energy Phys. {\bf 09} (2000) 028; 
J.A. Minahan and B. Zwiebach, ibid {\bf 09} (2000) 029.

\bibitem{ks1} A. Khare and A. Saxena, J. Math. Phys. {\bf 47} (2006) 
092902.

\bibitem{zhou} Y. Zhou, M. Wang, and Y. Wang, Phys. Lett. {\bf A 308} 
(2003) 31.

\end{thebibliography}
\end{document}